\documentclass[aps,prl,twocolumn,superscriptaddress,showpacs]{revtex4}
\usepackage{graphicx}
\usepackage{dcolumn}% Align table columns on decimal point
\usepackage{bm}% bold math
\begin{document}

%\preprint{}
\title{Linear magnetization dependence of the intrinsic anomalous Hall effect}

\author{Changgan Zeng}
\affiliation{Department of Physics and Astronomy, The University of Tennessee, Knoxville, Tennessee 37996, USA}
\author{Yugui Yao}
\affiliation{Institute of Physics, Chinese Academy of Sciences, Beijing 100080, China}
\author{Qian Niu}
\affiliation{Department of Physics, University of Texas, Austin 78712, USA}
\author{Hanno H. Weitering}
\affiliation{Department of Physics and Astronomy, The University of Tennessee, Knoxville, Tennessee 37996, USA}
\affiliation{Condensed Matter Sciences Division, Oak Ridge National Laboratory, Oak Ridge, TN 37831, USA}

\date{\today}

\begin{abstract}
The anomalous Hall effect is investigated experimentally and theoretically for ferromagnetic thin films of Mn$_5$Ge$_3$.  We have separated the intrinsic and extrinsic contributions to the experimental anomalous Hall effect, and calculated the intrinsic anomalous Hall conductivity from the Berry curvature of the Bloch states using first-principles methods. The intrinsic anomalous Hall conductivity depends linearly on the magnetization, which can be understood from the long wavelength fluctuations of the spin orientation at finite temperatures. The \textit{quantitative} agreement between theory and experiment is remarkably good, not only near 0 K, but also at finite temperatures, up to about $\sim$240 K (0.8 \textit{T$_C$})\end{abstract}


\maketitle


The appearance of a transverse voltage or electric field \textit{\textbf{E$_y$}} in a metal or semiconductor in response to a longitudinal electric current \textit{\textbf{j$_x$}} and perpendicular magnetic field \textbf{\textit{B}} is known as the Hall effect \cite{Chien,Hall,Smith}.  In nonmagnetic materials, this transverse voltage arises from a deflection of charge carriers by the Lorentz force \textit{\textbf{j$_x$}}$\times$\textbf{\textit{B}}, resulting in a Hall resistivity \textit{$\rho=E_y/j_x$} that is proportional to the magnetic field for weak fields. Only one year after his discovery of this ``ordinary Hall effect", Edwin Hall found that the Hall resistivity in ferromagnetic metals acquires an extra term which depends on the magnetization of the samples \cite{Hall}. Subsequent studies \cite{Smith} found that this ``anomalous" term is proportional to the spontaneous magnetization M. Empirically, one finds
\begin{equation}
\rho_H=\rho_{OH}+\rho_{AH}=R_0B+R_s4\pi M
\end{equation}
\noindent where $\rho_{OH}$ is the ordinary Hall resistivity due to the Lorentz force in a perpendicular magnetic field \textit{\textbf{B}}, $\rho_{AH}$ the anomalous Hall resistivity, \textit{R$_0$} the ordinary Hall coefficient, and \textit{R$_s$} the anomalous or spontaneous Hall coefficient.

Early theoretical interpretations of the anomalous Hall effect (AHE) pointed toward asymmetric scattering of the spin polarized charge carriers in the presence of spin-orbit coupling. Such scattering mechanisms can explain most of the qualitative features of the AHE observed in experiments, including the linear \cite{Smit} or quadratic \cite{Berger} correlation with the longitudinal resistivity $\rho_{xx}$, \textit{i.e.},
\begin{equation}
\rho_{AH}=a\rho_{xx}+b\rho_{xx}^2
\end{equation}
The two terms on the right are traditionally known as the skew scattering \cite{Smit} and side jump \cite{Berger} contributions, respectively. However, quantitative agreement between scattering theories and experiment remained largely unsettled, in part because the scattering potentials are unknown. In recent years, inspired by the new insight on Berry phase effects on Bloch electrons \cite{Sundaram}, a number of groups have evaluated the \textit{intrinsic} anomalous Hall conductivity (AHC) for ferromagnetic semiconductors \cite{Jungwirth}, transition metals \cite{Yao} and oxides \cite{Fang}, using first principles calculations. The intrinsic contribution can be quite large and for the first time, quantitative agreement between theory and experiment appears quite reasonable.  Interestingly, this intrinsic effect does not take into account scattering by impurities or phonons. It stems from spin-orbit coupling in the crystal band structure, a mechanism originally due to Karplus and Luttinger \cite{Karplus}.  

In this Letter, we examine the magnetization dependence of the AHE in light of the new theory. We have performed measurements on single-crystal films of ferromagnetic Mn$_5$Ge$_3$, extracted the intrinsic AHC, and found that this intrinsic conductivity is linear in the magnetization over a wide range of temperatures. This is puzzling. Although a linear dependence on magnetization comes naturally from the scattering mechanisms, the \textit{intrinsic} mechanism often yields strongly nonlinear behavior \cite{Jungwirth,Yao,Fang}.  We solve this puzzle by invoking the spin-fluctuation picture for finite temperature magnetism.  First principles calculations for the AHC based on this picture yield almost perfect comparison with experiment. 

Mn$_5$Ge$_3$(0001) thin films were grown on Ge(111) by solid-phase epitaxy, following the procedures in Ref. 11, and by co-deposition of Mn and Ge on Ge buffered GaAs(111). Both types of film exhibit very similar transport properties at low temperature but a GaAs substrate is desirable for transport measurements above $\sim$200 K because it minimizes the parallel conductance through the substrate. For the co-deposition experiment, we adopted the following procedure. A GaAs(111) substrate was annealed in ultrahigh vacuum at about 600$^{\circ}$C. A 35-nm-thick Ge buffer layer was subsequently grown on top of GaAs(111) at 400$^{\circ}$C. Next, we co-deposited Mn and Ge in a 5:3 atomic ratio onto the Ge(111)-c(2$\times$8) buffer layer substrate at 330$^{\circ}$C. The growth rate was $\sim$6 \AA/min, and the typical film thickness is 45 nm. Sharp reflection high energy electron diffraction (RHEED) patterns confirm the epitaxial growth reported in Ref. 11. Magnetic properties were measured with a superconducting quantum interference device magnetometer. The longitudinal resistivity was measured using the standard four-point-probe technique. Hall measurements were performed following the procedures descibed in Ref. 12.

\begin{figure}
\includegraphics{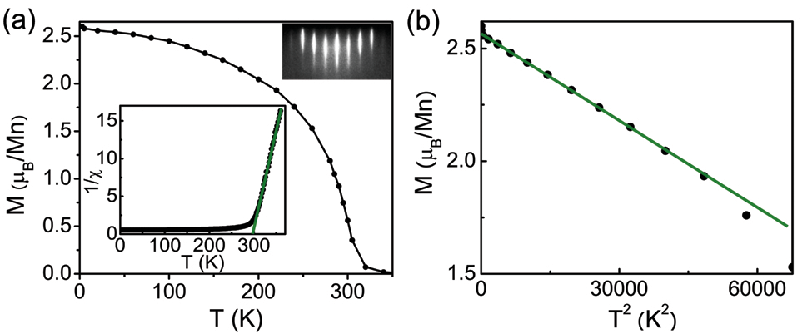}
\caption{\label{fig1}(Color online) (a) Temperature-dependent spontaneous magnetization \textit{M}. The upper inset shows the RHEED pattern of a Mn$_5$Ge$_3$ thin film grown on Ge/GaAs(111). The lower inset shows the temperature dependence of the inverse magnetic susceptibility, measured at 0.7 T. The green line is a linear fit and extrapolates to \textit{T$_C$} = 298 K. (b) \textit{M} versus \textit{T}$^2$. The green line is a linear fit, showing that $\Delta M(T)/M(0)\propto T^2$ up to $\sim$220 K.}
\end{figure}

Fig. 1 shows the spontaneous magnetization, obtained by extrapolating the high field part of the magnetization isotherms \textit{M(H)} to zero internal field, after correcting for the linear diamagnetic background signal. The saturation magnetization at 2.5 K is 2.6 $\pm$ 0.2 $\mu_B$ per Mn, in excellent agreement with our previous results from solid-phase epitaxy \cite{Zeng}. The susceptibility $\chi=M/B$ follows the Curie-Weiss law (Fig. 1(a): inset), and the Curie temperature \textit{T$_C$} determined from the linear fit is 298 $\pm$ 3 K. Our temperature range (2 - 400 K) is not low enough to resolve the T$^{3/2}$ dependence in the spontaneous magnetization expected from independent spin-wave excitations. Instead, we find a near perfect $T^2$ dependence up to about 220 K, as shown in Fig. 1(b). The $T^2$ fall of the magnetization arises from long wavelength, low frequency fluctuations of the magnetization, or multiple excitations of the interacting spin waves \cite{Lonzarich}.

Fig. 2 shows the Hall resistivity at various temperatures. The anomalous Hall component is well characterized by the ``knee" profile below \textit{T$_C$} and can still be identified at temperatures as low as 2.5 K (Fig. 2a). $R_0$ can in principle be extracted from the high field slope of the Hall isotherm \cite{Weitering}. The slope changes from negative to positive at about 180 K (Fig. 2(b)), indicating a sign change of $R_0$ \cite{Weitering}. $\rho_{AH}$ is obtained by extrapolating the high field slope to zero internal field, and is shown in Fig. 3(a), together with $\rho_{xx}$. The latter indicates a T$_C \approx$ 298 K, in excellent agreement with the magnetic results.

\begin{figure}
\includegraphics{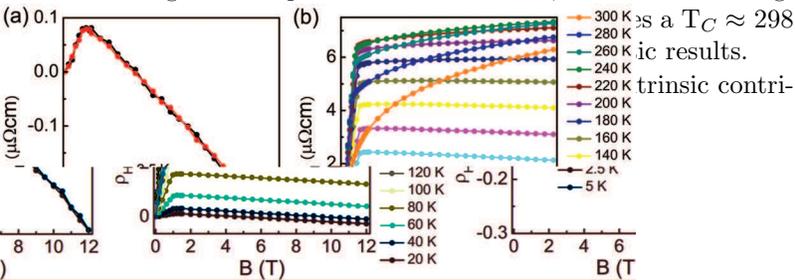}
\caption{\label{fig2}(Color online) Field-dependent Hall resistivity $\rho_H$ at various temperatures}
\end{figure}

The AHE contains both intrinsic and extrinsic contributions. In light of the semiclassical transport theory, the electrical current can be written as \cite{Yao} 
\begin{equation}
-\frac{e^2}{\hbar}\textbf{E}\times \int \frac{d^3k}{(2\pi)^3}\sum_{n}{f_n{\mathbf\Omega}_n(\textbf{k})}-\frac{e}{\hbar}\int \frac{d^3k}{(2\pi)^3}\sum_{n}{\delta f_n(\textbf{k})\frac{\partial \epsilon_n}{\partial \textbf{k}}}
\end{equation}                                                    
\noindent where ${\mathbf\Omega}_n$ is the Berry curvature of the Bloch state defined by ${\mathbf\Omega}_n(\textbf{k})=-$Im$\left\langle \nabla_\textbf{k}u_{n\textbf{k}}\left| \times  \right| \nabla_\textbf{k}u_{n\textbf{k}}\right\rangle$. The function $u_{n\textbf{k}}$ is the part of the Bloch wave function (with band index \textit{n}) that is periodic in the lattice. $f_n$ is the equilibrium Fermi-Dirac distribution function, and $\delta f_n$ is a shift proportional to the electric field and relaxation time. The first term is an intrinsic anomalous current and originates from the Berry curvature correction to the group velocity of a Bloch electron. This intrinsic contribution is independent of scattering and should lead to a quadratic dependence, $\rho_{AH} \propto \rho_{xx}^2$. The second term normally represents the longitudinal current but it will have a transverse component or Hall current in the presence of skew scattering. Because $\delta f_n$ is proportional to the relaxation time, the skew scattering contribution is proportional to $\rho_{xx}$. To separate the intrinsic and extrinsic contributions to the AHE, we thus write
\begin{equation}
\rho_{AH}=a(M)\rho_{xx}+b(M)\rho_{xx}^2
\end{equation}
where the functions $a(M)$ and $b(M)$ generally depend on the magnetization. In principle, it is impossible to uniquely separate the intrinsic and extrinsic contributions from temperature or field-dependent measurements on a single sample because temperature changes \textit{M}, $a(M)$, $b(M)$, and $\rho_{xx}$ simultaneously. However, the skew scattering contribution $a(M)$ is usually linear in magnetization \cite{Nozieres}. Accordingly, $a(M)$ can be obtained by plotting $\rho_{AH}/M\rho_{xx}$ versus $\rho_{xx}$, as shown in Fig. 3(b). This plot is linear below $\rho_{xx} \approx$ 110 $\mu \Omega$cm (or, equivalently $T \approx$ 220 K). The intrinsic contribution to the AHE is obtained by subtracting $a(M)\rho_{xx}$ from the experimental $\rho_{AH}$. The intrinsic anomalous Hall resistivity is also indicated in Fig. 3(a). The AHC is expressed as $\sigma_{AH}=\rho_{AH}/\rho_{xx}^2$ \cite{Lee}, so we immediately identify $b(M)$ as the intrinsic AHC, $\sigma_{IAH}$. The constant slope in Fig. 3(b) indicates that $b(M)/M$ is constant, hence the $\sigma_{IAH}$ is proportional to \textit{M}, \textit{i.e.},
\begin{equation}
\sigma_{IAH}\propto M
\end{equation}
This relation is the key result of our paper. Experimentally, this relation is valid between 1.7 and 2.6 $\mu_B$ as shown by the linear fit in Fig. 4(c). It spans an amazingly broad temperature interval of about 240 K or 0.8 \textit{T$_C$}. $\sigma_{IAH}$ extrapolates to 860 $\Omega^{-1}$cm$^{-1}$ at 0 K. Note that the uncorrected $\sigma_{AH}$ shows a non-monotonic dependence on \textit{M}.

\begin{figure}
\includegraphics{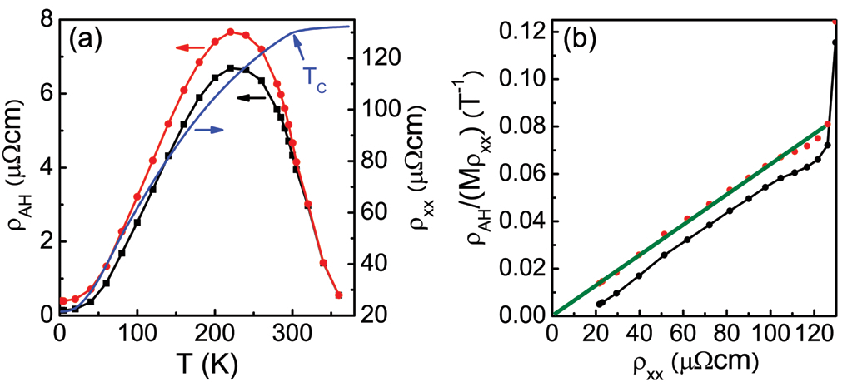}
\caption{\label{fig3}(Color online) (a) $\rho_{AH}$ as a function of temperature before (black line) and after (red line) subtracting the skew scattering contribution. $\rho_{xx}$ is also shown. (b) $\rho_{AH}/(M\rho_{xx})$ as a function of $\rho_{xx}$ before (black line) and after (red line) subtracting the skew scattering term. The green line is a guide to the eyes crossing the origin.}
\end{figure}

Based on the semi-classical transport theory \cite{Sundaram}, the intrinsic anomalous Hall conductivity (AHC) can be expressed as a sum of the Berry curvatures \cite{Yao}:
\begin{equation}
\sigma_{IAH}=-\frac{e^2}{\hbar}\int_{BZ}\frac{d^3k}{(2\pi)^3}{\mathbf\Omega}
^z(\textbf{k})
\end{equation}
\noindent where ${\mathbf\Omega}^z(\textbf{k})=\sum_{n}f_n{\mathbf\Omega}_n^z(\textbf{k})$. The intrinsic AHC can be evaluated from the electronic band structure. Following the procedures in Ref. 8, we first obtained a fully converged ground state by using the full-potential linearized augmented plane-wave method \cite{Blaha} with generalized gradient approximation \cite{Perdew} for the exchange-correlation potential. In this calculation, we sampled 1,000 \textbf{\textit{k}}-points in the first Brillouin zone, using K$_{max}$R$_{MT}$=9, where R$_{MT}$ and K$_{max}$ represent the muffin-tin radius and the maximum value of reciprocal-lattice vector, respectively. The muffin-tin radius of the Ge atom and Mn atom is 2.3 a.u.. In order to obtain even more accurate results, we also included semi-core \textit{3s}, \textit{3p} local orbitals for the Mn atom and \textit{3d} local orbitals for the Ge atom. Mn$_5$Ge$_3$ has a hexagonal D8$_8$-type crystal structure \cite{Forsyth}. We adopted the theoretical lattice constants, a=7.092 \AA\ and c=4.984 \AA, which are very close to the experimental values \cite{Zeng}. The internal parameters for the atomic positions, \textit{x}(Mn)=0.244 and \textit{x}(Ge)=0.606 also agree well with experimental data \cite{Forsyth}. The computed magnetization per unit cell is 26.5 $\mu_B$, in excellent agreement with the experimental magnetization. From the self-consistent potentials we obtain a fully converged $\sigma_{IAH}$ by summing the Berry curvature using a much larger set of \textbf{\textit{k}}-points. The final converged $\sigma_{IAH}$ value is 964 $\Omega^{-1}$cm$^{-1}$ at \textit{T} = 0 K, in excellent agreement with the experimental extrapolation of 860 $\Omega^{-1} $cm$^{-1}$.

Next, we explore various mechanisms that could lead to a linear magnetization dependence of $\sigma_{IAH}$ for $T >$ 0 K. A linear dependence on the magnetization is often observed and is automatically implied by treating spin orbit coupling as a linear perturbation in scattering theory \cite{Nozieres}. On the other hand, $\sigma_{IAH}$ of \textit{e.g.} Mn doped GaAs \cite{Jungwirth} and iron \cite{Yao} is nonlinear in spin-orbit coupling, and it appears difficult to obtain $\sigma_{IAH} \propto M$. More dramatically, in the case of SrRuO$_3$ Fang \textit{et al.} also found non-monotonic or even spiky dependences on the magnetization, which they attributed to ``magnetic monopoles" in momentum space \cite{Fang}.  To check whether $\sigma_{IAH}$ is linear in the spin-orbit coupling strength, we artificially changed the speed of light \textit{c}. This is equivalent to changing the spin-orbit coupling strength $\xi$ because $\xi \propto c^{-2}$ \cite{Yao}. Fig. 4(a) shows that $\sigma_{IAH}$ increases as $\xi$ increases, but it is clearly nonlinear in $\xi$. This nonlinearity means that spin-orbit interaction cannot be treated perturbatively for Mn$_5$Ge$_3$. It also implies that $\sigma_{IAH} \propto M$ cannot be obtained from a perturbative analysis of spin-orbit coupling. 

\begin{figure}
\includegraphics{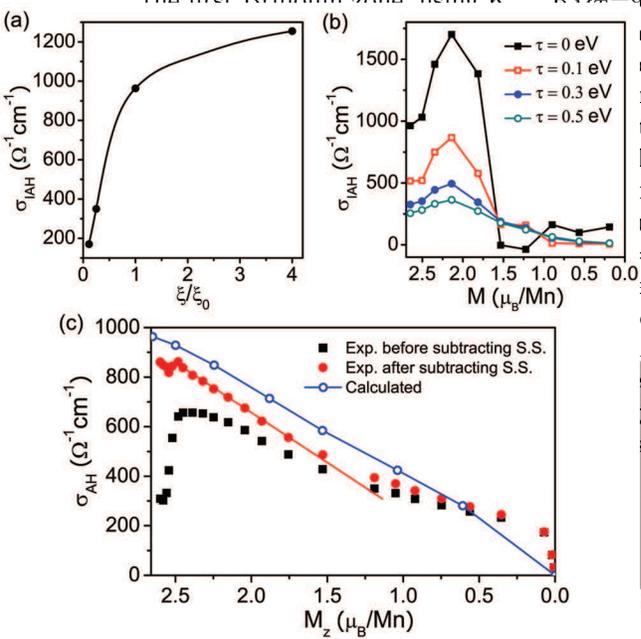}
\caption{\label{fig4}(Color online) (a) Calculated $\sigma_{IAH}$ versus the effective spin-orbit coupling strength $\xi$ relative to the real value $\xi_0$ for Mn$_5$Ge$_3$. (b) Calculated $\sigma_{IAH}$ versus the magnitude of magnetization for different relaxation time $\tau$. (c) Experimental AHC as a function of \textit{M} before (black dots) and after (red dots) subtracting the skew scattering (S.S.) contribution. The red line is linear fit. Calculated $\sigma_{IAH}$ values versus \textit{z}-component of the magnetization $M_z$ are connected by the blue line.}
\end{figure}

Next, we investigate whether $\sigma_{IAH} \propto M$ can be obtained by varying the spin-polarization or exchange splitting of the spin-up and spin-down bands of the Stoner model, as was done in Ref. 9. After obtaining self-consistent electronic charge density, we first tuned the exchange potential, then recalculated the wave functions and band structure (here no self-consistent calculations is needed and desired), and finally obtained $\sigma_{IAH}$. The result is shown in Fig. 4(b) for various relaxation times $\tau$. The qualitative and quantitative disagreement with experiment is evident. 

We now surmise that the linear dependence on the magnetization for $\sigma_{IAH}$, can be \textit{quantitatively} accounted for by the long-wavelength, low-frequency fluctuations of the spin orientation at finite temperatures. The experiment shows that $\sigma_{IAH}$ is linear in \textit{M} as long as the magnetization falls off quadratically (2 K $< T <$ 240 K). In this temperature interval, the magnitude of magnetization stays constant but the local magnetization rotates away from the \textit{z}-axis. We assume that the typical wavelength is much larger than the mean free path, so that we may employ the local approximation in the calculation of the conductivity tensor. In other words, we assume that the conductivity tensor is defined in each region and that the local conductivity can be calculated by considering an artificial infinite system with a uniform magnetization given by the local magnetization orientation in that region.  

Define the local spin orientation by the polar angles $(\theta,\varphi)$. The local intrinsic anomalous Hall conductivity is calculated as follows: (i) First, the self-consistent densities for both spin-up and spin-down electrons are obtained. (ii) Next, the spin quantization axis is rotated away from the \textit{z}-axis to a direction specified by $(\theta,\varphi)$. (iii) The Kohn-Sham orbitals are then recalculated in the presence of spin-orbit coupling, which are in the form of Bloch spinors. (iv) Finally, the Berry curvatures of the occupied Bloch spinors are calculated and summed over to obtain the intrinsic Hall conductivity, as was done in Ref.[8].

The observed intrinsic anomalous Hall conductivity should be a spatial average of the local values. The result is averaged over the azimuth angle $\varphi$ and found to be proportional to the \textit{z}-component of the local magnetization to a very good approximation. This is shown in Fig. 4(c) by the open circles. Because of this near-perfect linearity, we expect that further averaging over the small spread of the local \textit{z}-components will not change our result. The slope as well as numerical magnitude of the theoretical data are in almost perfect agreement with experiment, provided that the extrinsic skew scattering term is properly subtracted. The close agreement also suggests that a possible side jump contribution to the AHC (which also has $\rho_{AH} \propto \rho_{xx}^2$) must be very small or negligible. Theory and experiment deviate in the vicinity of \textit{T$_C$} where the interacting spin waves produce a different fall off of the magnetization.

In conclusion, the AHC of ferromagnetic Mn$_5$Ge$_3$ thin films has a large intrinsic contribution that varies linearly with the magnetization. The agreement between low temperature measurements and $T$ = 0 K \textit{ab initio} calculations is already compelling in its own right. Moreover, the temperature dependence of the intrinsic AHC can be \textit{quantitatively} accounted for by long wavelength fluctuations of the spin orientation up to 0.8 \textit{T$_C$} or 240 K.

We thank Dr. Steven C. Erwin for sharing the band structure results and Dr. Zhong Fang for assistance with the calculations and discussions. The experimental work was sponsored by NSF under Contract No. DMR 0306239 (FRG). The theoretical team (YGY and QN) also gratefully acknowledges support from DOE (DE-FG03-02ER45958), NSF (DMR-0404252), and NSFC of China (10404035, 10534030). Oak Ridge National Laboratory is managed by UT-Battelle, LLC, for the U.S. Department of Energy under Contract No. DE-AC05-00OR22725.

\end{document}